\documentclass[
    ,final            
  ]
  {aipproc}

\layoutstyle{6x9}

\newcommand{\bra}[1]{\langle {#1} |}
\newcommand{\ket}[1]{| {#1} \rangle}

\newcommand{\vecr}{{\mathbf r}}
\newcommand{\vecR}{{\mathbf R}}
\newcommand{\veck}{{\mathbf k}}

\begin{document}

\title{Continuum Response and Reaction in Neutron-Rich Be Nuclei}

\author{Takashi Nakatsukasa}{
  address={Department of Physics, Tohoku University, Sendai 980-8578, Japan}
}

\author{Manabu Ueda}{
  address={Akita National College of Technology, Akita 011-8511, Japan}
}

\author{Kazuhiro Yabana}{
  address={Institute of Physics, University of Tsukuba, Tsukuba 305-8571, Japan}
}

\begin{abstract}
We study $E1$ resonances, breakup and fusion reactions for
weakly bound Be nuclei.
The absorbing-boundary condition (ABC) is used to describe both
the outgoing and incoming boundary conditions.
The neutron continuum plays important roles in
response and reaction of neutron drip-line nuclei.
\end{abstract}

\maketitle


\section{Introduction}

At the drip line, since the separation energy becomes close to zero,
theoretical studies of nuclear structure and reaction
require continuum wave functions.
We have recently investigated an efficient and comprehensive method of
treating the continuum \cite{NY01,NY02-P1,NY02-P2,YUN02-P,UYN02}.
This is practically identical to the one called
``Absorbing Boundary Condition (ABC) method''
in the chemical reaction studies \cite{SM92}.
The method allows us to calculate the continuum wave functions
with the outgoing asymptotic behavior of many-body systems.

The essential trick for the treatment of the continuum in the ABC method is
to allow the infinitesimal imaginary part in the Green's function,
$i\epsilon$, to be a function of coordinate and finite, $i\epsilon(\vecr)$.
The $\epsilon(\vecr)$ should be zero in the interacting region and
be positive outside the physically relevant region of space.
Wave functions obtained using the ABC method are meaningful only in the
interacting region.
However, this provides all the necessary information
to obtain the scattering matrix.

In order to understand how the ABC method is functioning in later
applications, let us consider the potential scattering of a particle.
The scattering wave function is given by
\begin{equation}
\label{scattering_wave}
\ket{\psi_{\veck}^{(+)}} = \ket{\veck} + \frac{1}{E-H+i\epsilon}V\ket{\veck}
                         \equiv \ket{\veck} + \ket{\psi^{(+)}_{\rm scat}}.
\end{equation}
The scattering amplitude, $f(\Omega)$, is usually
defined by its asymptotic behavior
\begin{equation}
\psi_{\veck}^{(+)}(\vecr) \rightarrow \exp(i\veck\cdot\vecr)
                                   +f(\Omega)\frac{\exp(ikr)}{r} ,
\quad\quad (r\rightarrow\infty),
\end{equation}
but can be also written in a form
\begin{equation}
\label{f}
f(\Omega)=-\frac{m}{2\pi\hbar^2}\bra{\veck'}V\ket{\psi_{\veck}^{(+)}}
  =-\frac{m}{2\pi\hbar^2}
      \int d\vecr \exp(-i\veck'\cdot\vecr)V(\vecr)\psi_{\veck}^{(+)}(\vecr) ,
\end{equation}
where $\Omega$ is the direction of $\veck'$ and $|\veck'|=|\veck|$.
Equation \eqref{f} implies that the $f(\Omega)$ can be
determined by the scattering wave function,
$\psi^{(+)}_\veck(\vecr)$, in the interacting region where $V\neq 0$.
In other words, the $\psi^{(+)}_\veck(\vecr)$ outside the
interacting region is not needed to determine the scattering
properties.
This is why we are allowed to add the absorbing potential, $-i\epsilon(\vecr)$,
to the Hamiltonian as long as the $\psi^{(+)}_\veck(\vecr)$ in the
interacting region is correctly described.

In this paper, we present examples of the ABC method.
The method well works in the time-dependent and time-independent approaches
of mean-field and few-body models.

\section{Application of absorbing boundary condition (ABC)}

\subsection{Time-dependent-Hartree-Fock (TDHF) calculation
            in the 3D coordinate space}

In studies of giant resonances,
effects of the continuum has been treated in the random-phase
approximation (RPA) with Green's function in the coordinate space \cite{SB75}.
However, it is very difficult to directly apply the method to deformed nuclei
because construction of the Green's function becomes a difficult task
for the multi-dimensional space.
We have shown that the ABC method is very useful to treat
the electronic continuum in deformed systems,
such as molecules and clusters \cite{NY01}.
We have also investigated the applicability of the ABC in studies of
nuclear response calculations \cite{NY02-P1,NY02-P2}.
In this section, we discuss properties of the giant dipole resonance (GDR)
in nuclei at the $N=Z$ line, $^8$Be, and at the neutron drip line,
$^{14}$Be.

We use the ABC in the time-dependent Hartree-Fock (TDHF) calculations
on a three-dimensional (3D) coordinate grid.
The time evolution of the TDHF state, $\det\{\phi_i(\vecr,t)\}$,
is computed by
applying the electric dipole ($E1$) field to the Hartree-Fock (HF)
ground state:
\begin{eqnarray}
\label{TDHF}
i\frac{\partial}{\partial t}\phi_i(\vecr,t)
=\left( -\frac{1}{2m}\nabla^2
        + V_{\rm HF}(\vecr,t)
        + V_{\rm ext}(\vecr,t)
        - i\epsilon(\vecr)  \right) \phi_i(\vecr,t) ,
&& i=1\sim A,\\
\label{Vext}
V_{\rm ext}(\vecr,t) = k r_\alpha
  \left\{\frac{1}{2}\left( 1-\tau_z\right)e-\frac{Ze}{A}\right\}
  \delta(t) ,&& r_\alpha=x,y,z ,
\end{eqnarray}
where $k$ should be small enough to validate
the linear response approximation.
We calculate the expectation values of the $E1$ operator as a function
of time, then Fourier transforming to get the energy response
\cite{NY01,NY02-P1,NY02-P2}.
Since all frequencies are contained in the initial perturbation,
the entire energy response can be calculated with a single time evolution.

We use the Skyrme energy functional of EV8 \cite{BFH87} with
the SGII parameter set.
For the time evolution of the TDHF state,
we follow the standard prescription \cite{FKW78}.
The model space is a sphere whose radius is 22 fm.
The $i\epsilon(r)$ is zero in a region of $r<10$ fm,
while it is non-zero at $r>10$ fm.
The TDHF single-particle wave functions
are discretized on a rectangular mesh
in a 3D real space.
Now, we can solve Eq.~\eqref{TDHF} together with the
vanishing boundary condition at $r=22$ fm.
Time evolution is carried out up to $T=30\ \hbar$/MeV.

\begin{figure}[ht]
\includegraphics[height=0.3\textheight]{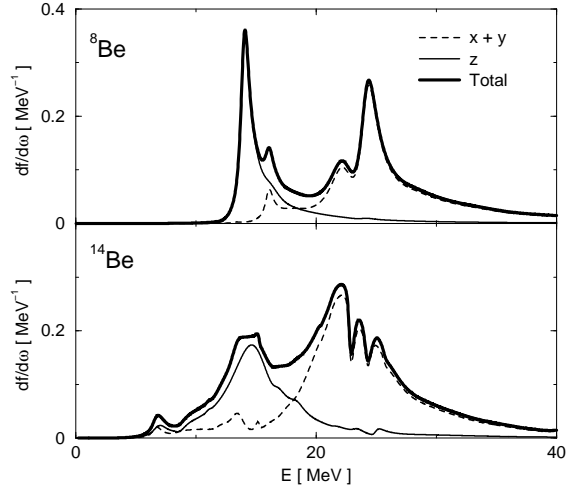}
\caption{Calculated $E1$ strength in $^{8,14}$Be.
Thin solid and dashed lines indicate the response to dipole fields
parallel and perpendicular to the symmetry axis, respectively.
Thick line shows the total strength.
}
\label{fig:Be8_14}
\end{figure}

The density distribution of the ground states of $^8$Be and $^{14}$Be
both possess a prolate superdeformed shape.
The $E1$ oscillator strengths are shown
in Fig.~\ref{fig:Be8_14}.
Here, we use a smoothing parameter of $\Gamma=0.5$ MeV.
The deformation splitting of the GDR peak is as large as about 10 MeV.
Since the ground-state shape is almost identical between the two isotopes,
average peak positions are similar.
However, the peak width is very different between the isotopes.
In $^{14}$Be at the neutron drip line,
the double-peak structure is almost smeared out in the total
strength (thick line).
On the other hand,
we observe prominent double-peak structure in $^8$Be.
The significant peak broadening in $^{14}$Be may attribute to
the small neutron separation energy and the Landau damping.

\subsection{Calculation of breakup reaction in the coordinate space}

Let us consider a reaction of a projectile, composed of 
core (C) plus neutron (n), on a target nucleus (T). 
Denoting the projectile-target relative coordinates by ${\bf R}$ and 
the neutron-core relative coordinates by ${\bf r}$, the Hamiltonian 
of this three-body system with the ABC is expressed as
\begin{equation}
\label{H}
H = -\frac{\hbar^2}{2\mu}\nabla_\vecR^2
      -\frac{\hbar^2}{2m} \nabla_\vecr^2
      +V_{nC}(\vecr) + V_{nT}(\vecr_{nT}) 
      +V_{CT}({\bf R}_{CT})
      -i\epsilon(\vecR,\vecr)
\end{equation}
where $\mu$ and $m$ are the reduced masses of projectile-target relative 
motion and neutron-core relative motion, respectively. $V_{nC}$,
$V_{nT}$, $V_{CT}$ are the interaction potentials of constituent
particles.
The $i\epsilon$ vanishes in the region
where the nuclear interactions are active.

The wave function may be expressed as a 
sum of the Coulomb wave, $ \psi^{(+)}({\bf R})$,
in the incident channel and the scattered wave.
\begin{equation}
\Psi^{(+)}({\bf R},{\bf r})
= \psi^{(+)}({\bf R}) \phi_0({\bf r}) + \Psi_{\rm scat}({\bf R},{\bf r})
\end{equation}
where $\phi_0({\bf r})$ is the ground state of the projectile,
described as a n-C bound state.
The $\Psi_{\rm scat}$ satisfies the
following inhomogeneous equation,
\begin{equation}
\left\{ E + e_0 -H \right\}
\Psi_{\rm scat}({\bf R},{\bf r})
=
\left\{ V_{nT}(\vecr_{nT})+V_{CT}(\vecR_{CT}) 
- V_C(\vecR) \right\}
\psi^{(+)}({\bf R}) \phi_0({\bf r}),
\label{3Bscat}
\end{equation}
where $V_C$ is the Coulomb distorting potential, $E$ is the bombarding
energy, and $e_0$ is the ground-state
energy of the projectile.
One should note that the right hand side of Eq.~\eqref{3Bscat}
is a localized function if we can neglect the difference between
$V_{CT}$ and $V_C$ at large $R$.
In \cite{UYN02},
we have studied a deuteron breakup reaction and compared our results
with those of the continuum discretized coupled channel (CDCC)
calculation.
Readers may refer to \cite{UYN02} for numerical details.
In this paper, we report the application to a nuclear breakup reaction of
$^{11}$Be on $^{12}$C.

The $^{10}$Be-n potential is taken as a Woods-Saxon shape whose depth
is set so as to produce the $2s$ orbital binding energy.
We adopt the optical potentials for $^{10}$Be-$^{12}$C
and for $n$-$^{12}$C.
The radial region up to 30 fm and 50 fm are used for $R$ and $r$, 
respectively.
The $i\epsilon_(R,r)$ is non-zero in the region of
20 fm $< R <$ 30 fm and in that of 25 fm $< r <$ 50 fm.
The n-$^{10}$Be relative angular momenta are included up to $l=3$.

\begin{figure}[ht]
\begin{minipage}[t]{0.4\textwidth}
\includegraphics[height=0.24\textheight]{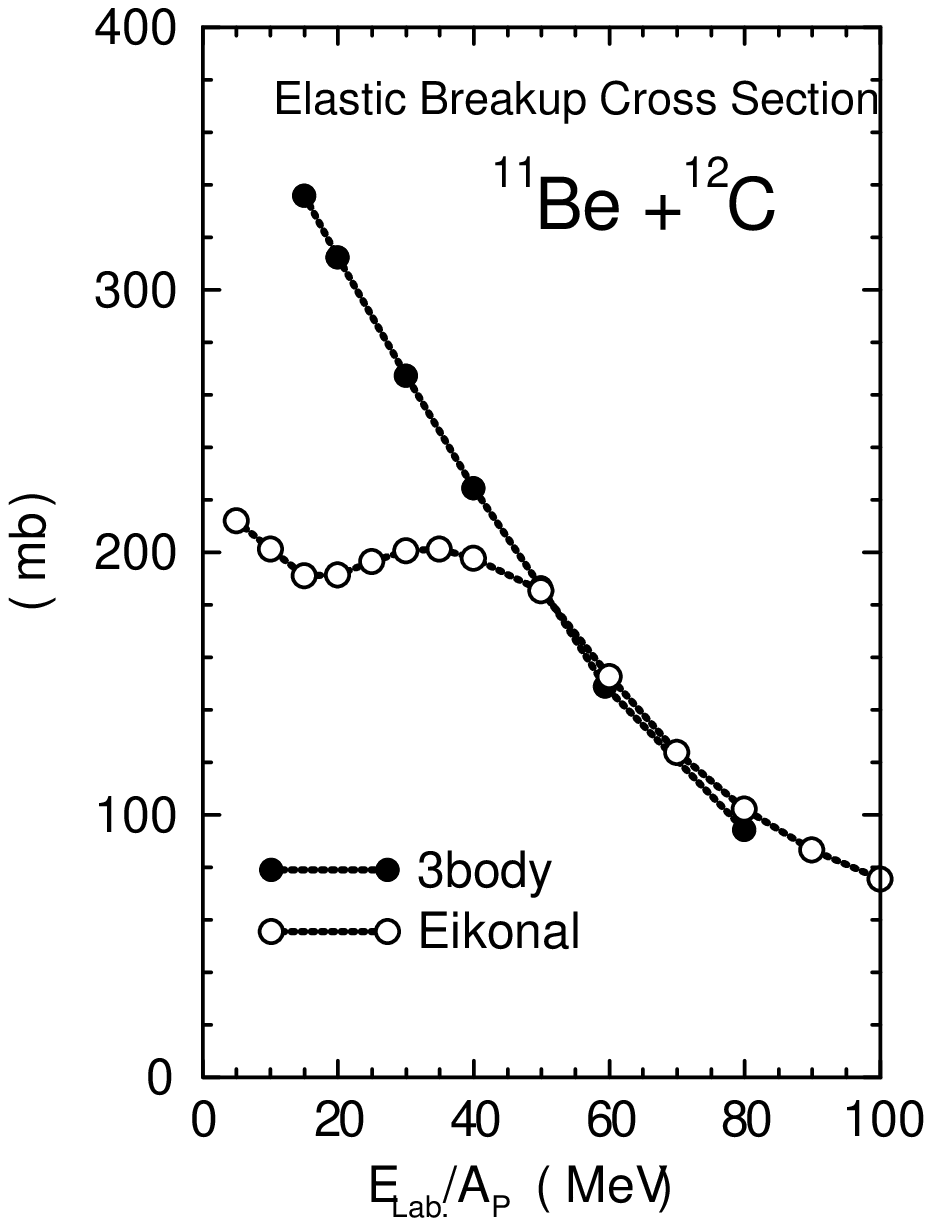}
\end{minipage}
\hfill
\begin{minipage}[t]{0.6\textwidth}
\includegraphics[height=0.2\textheight]{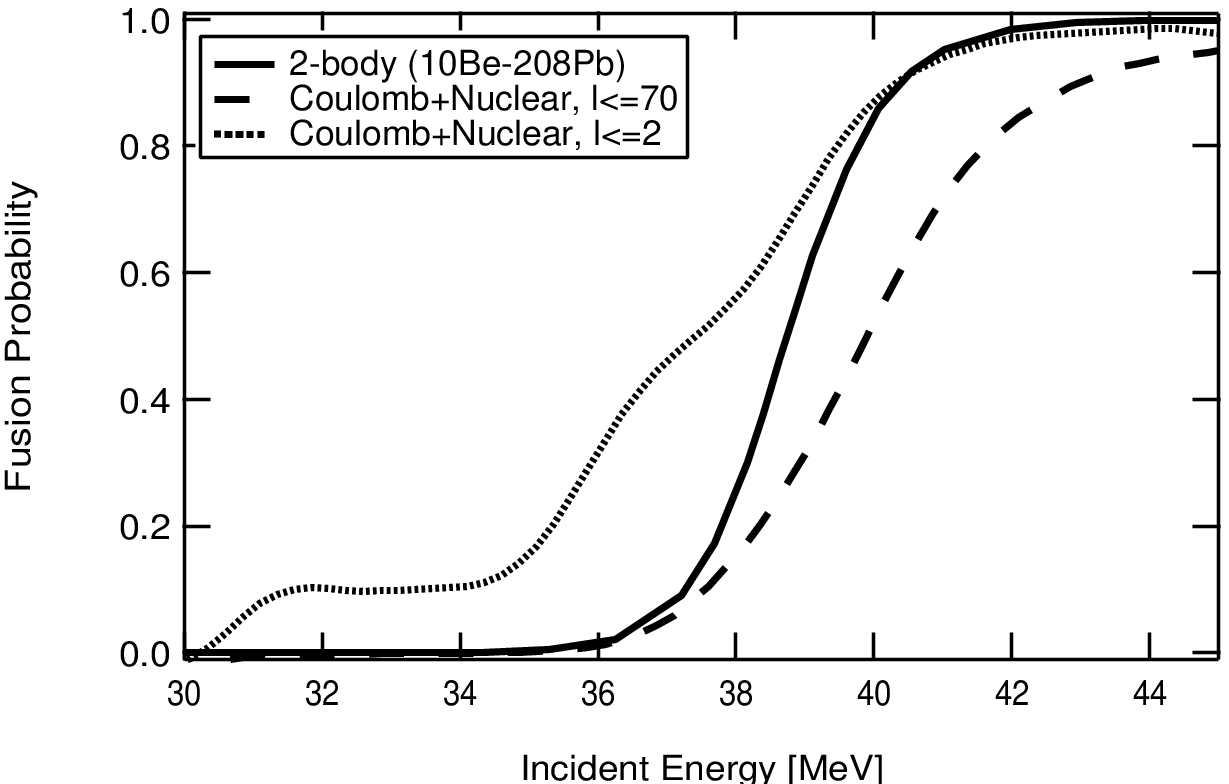}
\end{minipage}
\caption{({\bf Left})
The elastic breakup cross section in $^{11}$Be-$^{12}$C reaction.
Quantum calculation with ABC (closed circles)
is compared with the eikonal calculation (open).\newline
({\bf Right}) Calculated fusion probability
for $^{11}$Be$+^{208}$Pb as a function
of the projectile-target incident energy for head-on case ($J=0$).
Different truncation for n-C partial waves is taken for a dotted line
($l\leq 2$) and for a dashed line ($l\leq 70$).
The solid line is the fusion probability for $^{10}$Be$+^{208}$Pb
without a halo neutron.
}
\label{fig:11Be}
\end{figure}

In the left panel of Fig.~\ref{fig:11Be},
we show the elastic breakup cross sections 
of $^{11}$Be-$^{12}$C reaction.
The filled circles are the result of the ABC calculation
and the open circles for the eikonal calculation.
The elastic breakup cross section is substantially larger than that in the 
eikonal approximation at lower incident energy.
The failure of the eikonal approximation is apparent
at the incident energy below 50 MeV/A.
There, the quantum-mechanical treatment is required for the
three-body continuum.

\medskip

\subsection{Time-dependent-wave-packet calculation for fusion probability}

So far, we have used the ABC to simulate the outgoing boundary condition,
$i\epsilon$ being non-zero outside the interacting region.
In this section, we set the $i\epsilon$ non-zero
inside the Coulomb barrier,
to simulate the incoming boundary condition.

We consider a fusion reaction of $^{11}$Be on $^{208}$Pb near the Coulomb
barrier energy.
Again, $^{11}$Be is described as a weakly-bound system of a neutron
and $^{10}$Be core.
The target is now $^{208}$Pb.
All the nuclear potentials are taken to be real and of the Woods-Saxon type.
The $i\epsilon(R_{CT})$ is also of the Woods-Saxon shape and
corresponds to the imaginary part of
the C-T optical potential which describes the fusion between
core and target.
See \cite{YUN-u} for details.
The time-dependent Schr\"odinger equation,
\begin{equation}
i\frac{\partial}{\partial t} \Psi(\vecR,\vecr,t)=H\Psi(\vecR,\vecr,t),
\end{equation}
where $H$ is given by Eq.~\eqref{H},
is solved with an initial
wave function
\begin{equation}
\Psi(\vecR,\vecr,t=0)=\exp(-iKR)\Phi_0(\vecR)\phi_0(\vecr) ,
\end{equation}
where $\Phi_0(\vecR)$ is a Gaussian wave packet.
The initial wave packet, $\Phi_0(\vecR)\phi_0(\vecr)$,
exists in a region beyond the range of
nuclear interaction of $V_{nT}$ and $V_{CT}$.
The wave packet is going to approach to the target, then
if the $^{10}$Be core penetrates the Coulomb barrier,
the wave function disappears because of absorption by
the imaginary potential $i\epsilon(R_{CT})$.
The ratio of the flux loss against the initial flux gives
the fusion probability.
In this calculation, since the final destination of neutron is irrelevant,
the calculated fusion probability contains
both complete and incomplete fusions.

The energy projection \cite{YUN-u} leads to the fusion probability
shown in Fig.~\ref{fig:11Be} (right).
When we include the n-C partial angular momenta only up to $l=2$, the result
indicates a strong fusion enhancement at sub-barrier energies, compared to
the two-body fusion calculation of $^{10}$Be and $^{208}$Pb.
However, when we enlarge our model space to include $l\leq 70$,
this enhancement disappears and we even have suppression.
This is consistent with similar study for fusion suppression obtained
for $^{11}$Be and $^{40}$Ca \cite{Yab97}.
The results suggest necessity of high partial waves for the n-C relative
motion to describe correct dynamics of neutron breakup and transfer
under the strong Coulomb field.





\bibliographystyle{aipproc}   

\bibliography{myself,nuclear_physics,chemical_physics}

\IfFileExists{\jobname.bbl}{}
 {\typeout{}
  \typeout{******************************************}
  \typeout{** Please run "bibtex \jobname" to optain}
  \typeout{** the bibliography and then re-run LaTeX}
  \typeout{** twice to fix the references!}
  \typeout{******************************************}
  \typeout{}
 }

\end{document}